\documentclass[twoside]{article}
\usepackage{epsf}
\usepackage{cite}
\usepackage{subeqnarray}
\usepackage{fleqn}
\usepackage{espcrc2}

\newcommand{\rf}[1]{(\ref{#1})}
\newcommand{\bea}{\begin{eqnarray}}
\newcommand{\eea}{\end{eqnarray}}
\newcommand{\e}{{\rm e}}

\renewcommand{\b}{\beta}

\newcommand{\del}{\delta}
\newcommand{\D}{\Delta}
\renewcommand{\L}{\Lambda}

\newcommand{\cD}{{\cal D}}

\def\void{}
\def\labelmark{}

\newenvironment{formula}[1]{\def\labelname{#1}
\ifx\void\labelname\def\junk{\begin{displaymath}}
\else\def\junk{\begin{equation}\label{\labelname}}\fi\junk}%
{\ifx\void\labelname\def\junk{\end{displaymath}}
\else\def\junk{\end{equation}}\fi\junk\labelmark\def\labelname{}}

{\ifx\void\labelname\def\junk{\end{array}\end{displaymath}}
\else\def\junk{\end{array}\right.\end{equation}}
\fi\junk\labelmark\def\labelname{}\def\junk{}
}

\newcommand{\beq}{\begin{formula}}
\newcommand{\eeq}{\end{formula}}
\newcommand{\beqv}{\begin{formula}{}}
\newcommand{\bes}{\begin{subeqnarray}}
\newcommand{\ees}{\end{subeqnarray}}
\newcommand{\beqa}{\begin{eqnarray}}
\newcommand{\eeqa}{\end{eqnarray}}

\title{Spin-spin correlation functions of spin systems coupled to 2-d quantum
gravity for $0 < c < 1$. }

\author{J. Ambj\o rn and K. N. Anagnostopoulos\address{The Niels Bohr 
Institute\\
Blegdamsvej 17, DK-2100 Copenhagen \O , Denmark}, %
U. Magnea\address{Nordita,\\
 Blegdamsvej 17, DK-2100 Copenhagen \O , Denmark}%
and G. Thorleifsson\address{Physics Department, Syracuse University,\\
 Syracuse, NY 13244, USA}}

\begin{document}

\begin{abstract}
We perform Monte Carlo simulations of 2-d dynamically triangulated
surfaces coupled to Ising and three--states Potts model matter. By
measuring spin-spin correlation functions as a function of the
geodesic distance we provide substantial evidence for a diverging
correlation length at $\beta_c$. The corresponding scaling exponents
are directly related to the KPZ exponents of the matter fields as
conjectured in \cite{ajw}.
\end{abstract}

\maketitle

\section{INTRODUCTION}
The calculation of the dressed scaling exponents
is a milestone in the theory of two-dimensional quantum gravity
\cite{kpz}. Strictly speaking, however,  the derivation uses only
finite size scaling arguments and involves only matter field
correlators integrated over all space--time. The concept of a
diverging correlation length at the critical point associated with the
correlators is not used in the derivation and was never shown to
actually {\it exist}. It is possible to define a correlation length by
defining the two point correlator of a field $\phi(x)$ in terms of the
reparametrization invariant geodesic distance $R$ between two marked
points:
\beqa
& G_{\phi} (R;\L) = \int \cD [g] \cD [\phi] \;\e^{-S_G-S_M} 
                        \int d^2x d^2y \nonumber\\ 
& \quad\times\sqrt{g(x)g(y)} \; 
 \phi(x) \phi(y) \del( D_g(x,y) -R).
\eeqa
Even with this definition at hand, it is possible that the correlation
length of the system coupled to gravity does not diverge, even though
it does in flat space and the transition when coupled to gravity is
continuous \cite{ah}.  The fixed volume correlator $G_{\phi}(R;V)$ is
obtained by Laplace transforming $G_{\phi}(R;\L)$ and from it one
obtains the correlators $S_\phi(R;V)\equiv G_{\phi} (R;V) / V \,
Z(V)$, where $Z(V)$ is the fixed volume partition function. The case
$\phi=1$ corresponds to the volume--volume correlators $G_{1}(R;V)$,
$G_{1}(R;\L)$ and the average ``area'' of a spherical shell $S_1(R;V)$
or radius $R$. In the case of a matter field $\phi$ with scaling
dimension $\D_0$ in flat space, $\int_0^\infty dR \; S_\phi(R;V) \sim
V^{1-\D}$ where $\D$ is the dressed scaling dimension of the
field. From this scaling and the corresponding behaviour of
$S_\phi^{(0)}(R;V)$ in flat space
\beq{*12}
S^{(0)}_\phi(R;V) \sim  \frac{R}{R^{2\D_0}} f(x) =
V^{1/2-\D_0} F^{Flat}_\phi(x) ,
\eeq
we may conjecture that \cite{ajw} 
\beq{*11}
S_\phi(R; V) \sim \frac{R^{d_h-1}}{ R^{d_h \D}} f(x) =
V^{1-\D-1/d_h} F_\phi(x),
\eeq
where $f(0) >0$, $F(x)\sim x^{d_h(1-\D)-1}$ for small
$x=R/V^{1/d_h}$ and $d_h$ is the fractal dimension of space--time as
defined for example by the finite size scaling relation $S_1(R;V)\sim
V^{1-1/d_h}F_1(x)$.  In this article we report on extensive numerical
simulations which provide substantial evidence that Eq.~\rf{*11} holds
for matter fields with $0<c<1$. This points to the existence of a
diverging correlation length of the conformal field theory coupled to
gravity if we use the geodesic distance as a measure of length.

\section{SIMULATIONS AND RESULTS}
\begin{figure}[t]
\centerline{\epsfxsize=3in \epsfysize=2in \epsfbox{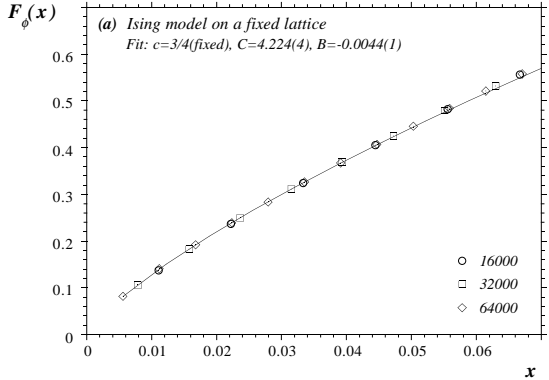}}
\caption{
Data for $F^{Flat}_\phi(x)$ as defined in
Eq.\rf{*12} for small values of $x=r/N^{1/d_h}$, $d_h=2$. 
The fit is to Eq.\rf{*27}. 
}
\label{f:1}
\end{figure}

\begin{figure}[t]
\centerline{\epsfxsize=3in \epsfysize=2in \epsfbox{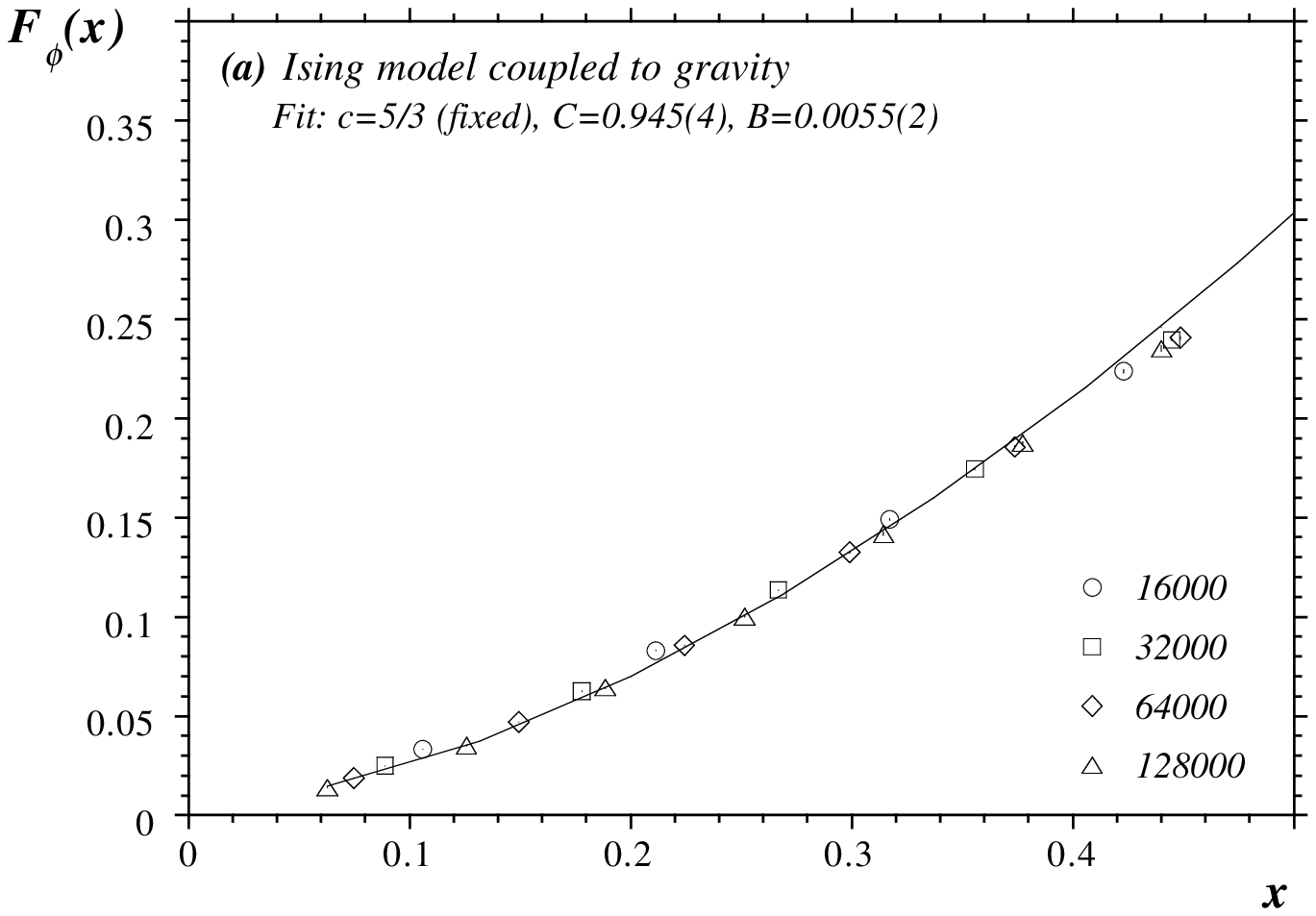}}
\centerline{\epsfxsize=3in \epsfysize=2in \epsfbox{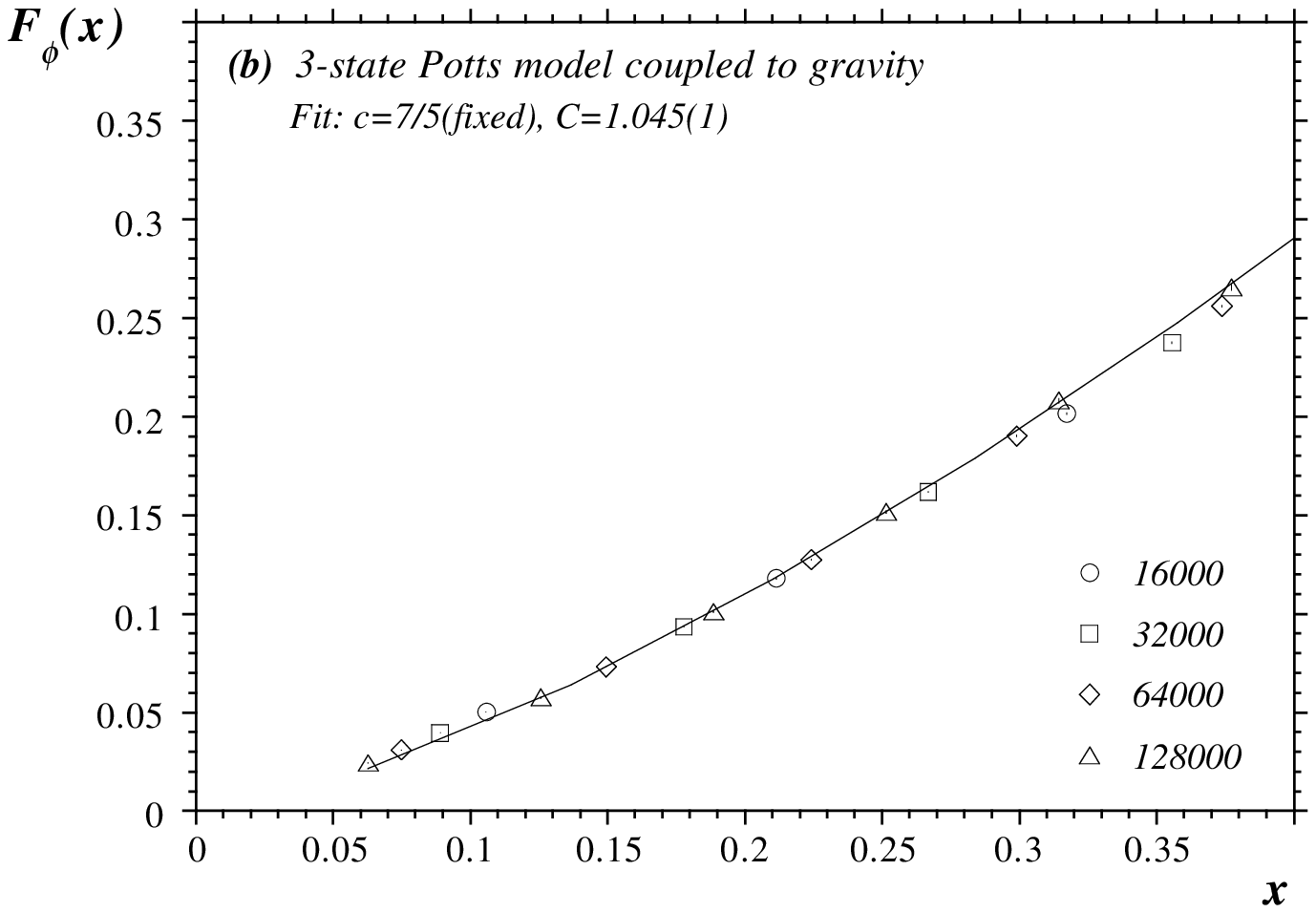}}
\caption{
({\it a}) Same as in Fig.~\ref{f:1} 
for $F^{Ising}_\phi(x)$ for
$d_h=4$. ({\it b}) Same as in ({\it a}) for $F^{Potts}_\phi(x)$.
}
\label{f:2}
\end{figure}

We perform numerical simulations on dynamically triangulated surfaces
with $S^2$ topology with tadpoles and self-energy diagrams. We place
Ising and three--states Potts model spins on the vertices of the surface
which interact with the spins of the neighbouring
vertices. The Monte Carlo updating of the triangulations is performed
by the so{--}called flip algorithm and the spins are updated by
standard cluster algorithms. The flips are organised in ``sweeps''
which consist of approximately $N_L$ {\it accepted} flips where $N_L$
is the number of links of the triangulated surface.  After a sweep we
update the spin system \cite{bj}.
The results presented in this paper cover system sizes from 16000 to
128000 triangles and the number of sweeps is $1.7${--}$5.0\times
10^6$. The simulations are performed at the analytically known infinite
volume $\b_c$ \cite{bk}.
Geodesic distances $r$ on the triangulations are defined  as the
shortest link distance between two vertices. The discretized volume of
the system is the number of triangles $N_T$ and we use the scaled
distance $x=R/N^{1/d_h}$, where $N$ is the number of vertices. We
measure the discretized distributions corresponding to $S_{1}(R;V)$
and $S_{\phi}(R;V)$ 
\bes
\label{*21all}
\slabel{*21}
n_1    (r;N) &=& \langle \sum_j \,\delta(D_{ij}-r) \rangle\, ,\\
\slabel{*21a}
n_\phi (r;N) &=& \langle \sum_j \,\sigma_i\sigma_j\,\delta(D_{ij}-r) \rangle
   \, ,
\ees
where the indices $i$ and $j$ label vertices: $i$ is a random fixed
vertex for each measurement and $j$ runs over all vertices of the
given configuration.  $D_{ij}$ is the link distance between the
vertices labelled by $i$ and $j$. We determine the fractal dimension
$d_h$ by using $n_1(r;N)=N^{1-1/d_h}F(x)$ and $n_\phi(r;N)=N^{1-\D-1/d_h}
F_\phi(x)$. Our data is in very good agreement with the results
obtained in \cite{syracuse,ajw}, namely $d_h=4$ for both the Ising
model and the three-states Potts model coupled to gravity. 

\begin{table}[t]
\label{t:2}
\begin{center}
\caption{The parameters of the fits to Eq.~\rf{*27all} for the Ising 
 model on a flat $T^2$ for $N_T = 64000$ ($\D_0=1/8$), and for the Ising
model ($\D=1/3$) and the three--states Potts model ($\D=2/5$) on $S^2$
coupled to quantum gravity for $N_T = 128000$.}
\begin{tabular}{ c r r r  }
\hline
            &             &            &           \\
            & $\D_0=1/8$  &  $\D=1/3$  & $\D=2/5$  \\
            &             &            &           \\
\hline
            &             &            &           \\
$C_1      $ & 4.224(4)    &  0.945(4)  & 1.044(4)  \\
$B_1      $ & -0.0044(1)  &  0.0054(2) & 0.0003(3) \\
            &             &            &           \\
$c        $ & 0.765(1)    &  1.535(6)  & 1.396(5)  \\
$C_2      $ &  4.39(2)    &  0.834(7)  & 1.040(8)  \\
            &             &            &           \\
$c        $ &   0.753(4)  &  1.596(6)  & 1.470(8)  \\
$C_3      $ & 4.25(4)     &  0.870(7)  & 1.10(1)   \\
$B_3      $ &  -0.0038(7) &  0.0036(1) & 0.0055(3) \\
            &             &            &           \\
\hline
\end{tabular}
\end{center}
\end{table}

We test the hypothesis of Eq.~\rf{*11} by using the values of the
dressed scaling dimensions of the Ising and three--states Potts model
spin fields $\D=1/3,2/5$. The prediction for $d_h=4$ is
\bes
\label{*24all}
\slabel{*24}
F_\phi^{Ising}(x)  &\propto & x^{5/3}~~~~{\rm for}~~~x \ll 1\, , \\
\slabel{*25}
F_\phi^{Potts}(x)  &\propto & x^{7/5}~~~~{\rm for}~~~x \ll 1\, .
\ees
In order to calibrate the expected accuracy with which one can
determine the exponent $d_h(1-\D)-1$ we have performed simulations of
an Ising model on a fixed triangular lattice with periodic boundary
conditions ($\D_0=1/8$). We measured $F^{Flat}_\phi(x)$ of
Eq.~\rf{*12} and we expect that $F^{Flat}_\phi(x)\propto x^{3/4}$ for
$x \ll 1$.

In Table~\ref{t:2} we show the results of the fits for the largest
lattice to the following functional forms:
\bes
\label{*27all}
\slabel{*27}
F_\phi (x)        &=&  C_1 x^{d_h(1-\D)-1} +B_1 \, ,\\
\slabel{*27a}
                  &=&  C_2 x^{c} \, ,\\
\slabel{*27b}
                  &=&  C_3 x^{c} + B_3\, .
\ees
Our results are consistent with the exponents conjectured in
Eqs.~\rf{*24} and \rf{*25}.  We find that simulating the largest
lattices, 128000 triangles, is important for obtaining enough data
points in the relevant region $x < 0.45$.  The fits to the predicted
behaviour Eq.~\rf{*27} are good and the finite size correction $B$
approaches zero as the volume increases.  Even if the exponent $c$ is
allowed to vary, as in the fits to Eqs.~\rf{*27a} and \rf{*27b}, it
approaches its predicted value convincingly as the volume is
increased. The small discrepancy is consistent with finite size
effects, which clearly are more important than the statistical errors
quoted in Table~\ref{t:2}. 
We find that the difference
in the values of $c$ obtained from the fits to Eqs.~\rf{*27a} and
\rf{*27b} gives a measure of the systematic errors entering from
varying the range and type of the fits. 
For a more complete presentation of the data 
we refer the reader to \cite{aamt}.

\end{document}